\documentclass[twocolumn,superscriptaddress,showpacs]{revtex4}
\usepackage{epsfig}
\usepackage[centertags]{amsmath}
\usepackage{graphics}

\begin{document}
\title{Resonance width oscillation in the bi-ripple ballistic electron waveguide}
\author{Hoshik Lee}
\affiliation{ The Center for Complex Quantum Systems, The University
of Texas at Austin, Austin, Texas 78712}
\affiliation{Department of Physics, Temple University, Philadelphia, Pennsylvania 19122}
\author{L. E. Reichl}
\affiliation{ The Center for Complex Quantum Systems, The University
of Texas at Austin, Austin, Texas 78712}
\date{\today }

\begin{abstract}
Interference of quasi bound states is studied in a ballistic electron ripple waveguide with two ripple cavities whose distance apart can be varied. This system is the waveguide analog of Dicke's model for two interacting atoms in a radiation field.  Dicke's model has  resonances whose widths  change in an oscillatory manner as the distance between the  atoms is varied. Resonances that form in a bi-ripple waveguide behave in a manner that has some similarity to Dicke's system, but also important differences.  We numerically investigate the behavior of resonance widths in the waveguide as the distance between the two ripple cavities changes and we find that the resonance  widths oscillate with variation of distance, but the coupling does not decrease as it does in Dicke's system. We discuss  differences between our waveguide system and other systems showing the analogous of Dicke effect. We also study S-matrix pole trajectories and find that they rotate in counterclockwise direction on a circle in the complex energy plane.  
\end{abstract}

\preprint{CQS-3}

\pacs{72.10.-d, 42.50.Gy, 73.63.Kv, 73.23.Ad }

\maketitle

%
%
%
%
Interference is a distinctive property of waves in both quantum and classical mechanics. It plays an especially  important role in quantum mechanics because it demonstrates the wave nature of  particles.  Matter-wave interference has been observed in atomic, nuclear, and solid state systems.  However, these systems are often too difficult to control to allow  investigation of  a variety of interference phenomena. Recent developments of nano technology now enable us to study interference phenomena using ballistic electron waveguides, carbon nanotubes, quantum dots.  Indeed, quantum dots are often called "artificial atoms" because they have discrete energy levels due to the confinement of the electron wavefunction. One can change the electron state inside a quantum dot with external controls such as gate voltages. The Coulomb blockade conductance profile shows the discreteness of energy levels inside a quantum dot.  

Ballistic electron waveguides have  been used to study the interaction of  quantum systems with their environment. They have provided  an important tool for studying the effects of quantum chaos on scattering phenomena~\cite{Akguc:2001p185, Akguc:2003p193, Dembowski:2004p2213, Weingartner:2005p170, Rotter:2007p189, Lee:2006p1023}. In  waveguides, Coulomb blockade does not occur because discrete energy levels don't  exist. However, even though there is no discrete energy spectrum, quasi bound states can build up in waveguides and significantly affect the electron transmission. Since there exist both non-resonant  and resonant channels for  electron transmission, resonances  show the Fano profile, indicating the interaction between resonant and non-resonant channels. 

Dicke proposed that spontaneous radiation from two interacting atoms  would have a much longer wavelength than the distance between the  two atoms ~\cite{Dicke:1953p13582}. This  radiation is associated with a transition from a symmetric collective state of the atoms  to the ground state.  It  is called {\it superradiance}, because the symmetric state is strongly coupled to the external photon fields and its decay occurs more rapidly than the decay of excited states of independent atoms.  There is also an antisymmetic collective state which is  weakly coupled to the photon fields. Its decay  is called {\it subradiance} because it decays more slowly than independent atoms. The different coupling to the photon fields results in broadening and narrowing of the resonance widths, which is called {\it Dicke effect}. The analogous effect can be found in mesoscopic systems because quantum dots or quasi bound states in a waveguide  behave like atoms.  The Dicke effect in  mesoscopic systems was first found by Shahbazyan and Raikh ~\cite{Shahbazyan:1994p1192}  for  two channel resonant tunneling in a system with two impurities.   In their system, the bound state of each impurity are indirectly coupled through the electron wavefunction in external leads. The analogy to Dicke effect has since  been found in several other mesoscopic systems,  such as  quantum dots coupled via a  common phonon field~\cite{Vorrath:2003p991} and a quantum wire with side coupled quantum dots~\cite{Orellana:2006p1062}. The  Dicke effect has also been seen in the  spontaneous radiation from collectively interacting quantum dots  in an experiment by M. Scheibner {\it et. al.} in 2007~\cite{Scheibner:2007p500}. 

In a previous study~\cite{ Lee:2008p840}, we found the analog of the  Dicke effect in a multi-ripple ballistic electron  waveguide. Qusai bound states in each  cavity of the waveguide  interact  through the internal leads to form the collective states which result in  different coupling to the external  leads. We found the broadening and narrowing of resonance widths in electron transmission (conductance) as more ripple cavities are added to the waveguide. However, we could not determine how the resonance widths change as the distance between ripple cavities  varies,  which is one of the key features of the Dicke effect. In this brief report, we study the  dependence of resonance widths on the distance between the two cavities of a bi-ripple waveguide  the  Generalized Scattering Matrix Method(GSM)~\cite{TakSumChu:1986p776,Cahay:1988p816}. 

\begin{figure}
\begin{center}
\includegraphics[width=0.5\textwidth]{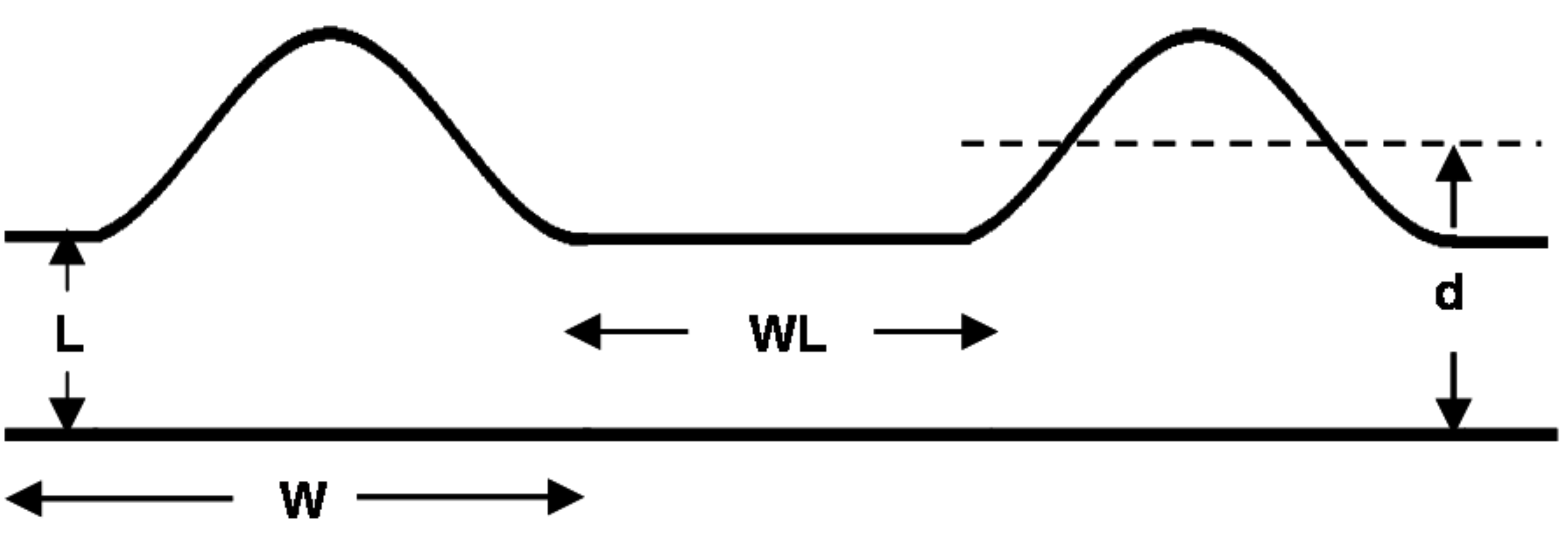}
\end{center}
\caption{\label{fig1} A bi-ripple waveguide with a flat waveguide between the two ripple cavities.}
\end{figure}

We consider a waveguide system in which  two ripple cavities are connected to each other by  a flat waveguide whose length is $WL$. The length of each  ripple cavity  is $W$, and the upper wall of each cavity  is described by an analytic function, $y=d-a{\rm cos}(2 \pi x/W)$ (See Fig.~\ref{fig1}). The outer ends of each cavity is connected to a semi-infinite lead  whose height is $L=(d-a)$. We use R-matrix theory to calculate the S-matrix for a single ripple cavity  ~\cite{Lee:2008p840}, \cite{Lee:2006p1023}.  R-matrix theory was originally developed by Wigner and Eisenbud in 1950s for the study of nuclear scattering. Recently, R-matrix theory has been used to study electron transmission in ballistic electron  waveguides  ~\cite{Akguc:2001p185, Akguc:2003p193, Lee:2006p1023, Lee:2008p840}. We consider a waveguide built in a 2DEG (two dimensional electron gas) made of a GaAs/AlGaAs heterostructure at a very low temperature. We then use the GaAs  effective mass of the electron $m^*=0.061 m_e$, where $m_e$ is the free electron mass. We use parameters $a=$ 13.846 ~\AA, $d=$ 47.269 ~\AA, and $W=$ 300 ~{\AA~}    for the waveguide in Fig.~\ref{fig1}. $E_1={\hbar^2 \pi^2\over 2m^* L^2}=0.503 4074 eV$ is the threshold energy to open the first propagating mode in the semi-infinite waveguide. We use $E_1$ as the unit of energy through this report. We only  consider the incident electron energies $E$  which allow one  propagating mode   so that $E_1 \leq E \leq 4E_1$. 

Since electrons freely propagate in the flat waveguide between the ripple cavities, the electron wavefunction acquires the phase factor $s_{WL}=e^{i k_1 WL}$, where $k_1=\sqrt{2m^*/\hbar^2(E-E_1)}$. The S-matrix for phase acquisition is a diagonal matrix, 
\begin{equation} S_{WL} \,=\, \left ( 
\begin{array}{cc}
s_{WL} & 0 \\
0 & s_{WL}
\end{array} \right).
\end{equation}
For a single ripple waveguide, the S-matrix is written in terms of transmission and reflection coefficients such as
\begin{equation}
S_1 \,=\, \left ( 
\begin{array}{cc}
r_1 & t'_1 \\
t_1 & r'_1
\end{array} \right).
\end{equation} 
The S-matrix for the second ripple waveguide $S_2$ can be obtained in the same way. If two ripple waveguides are identical, $S_1$ and $S_2$ are same. We consider only identical ripples in this report. The overall scattering matrix $S_T$ is obtained by the GSM~\cite{TakSumChu:1986p776,Cahay:1988p816},
\begin{equation}
S_T \,=\, \left(
\begin{array}{cc}
r_1 + t'_1 s_{WL} U_2 r_2 s_{WL} t_1 & t'_1 s_{WL} U_2 r'_2 \\
t_2 s_{WL} U_1 t_1 & r'_2+t_2 s_{WL} U_1 r'_1 s_{WL} t'_2
\end{array} \right ),
\end{equation}
where $U_1 \,=\, (1 \,-\, r'_1 s_{WL} r_2 s_{WL})^{-1}$ and $U_2 \,=\, (1 \,-\, r'_1 s_{WL} r_2 s_{WL})^{-1}$.

\begin{figure}
\begin{center}
\includegraphics[width=0.5\textwidth]{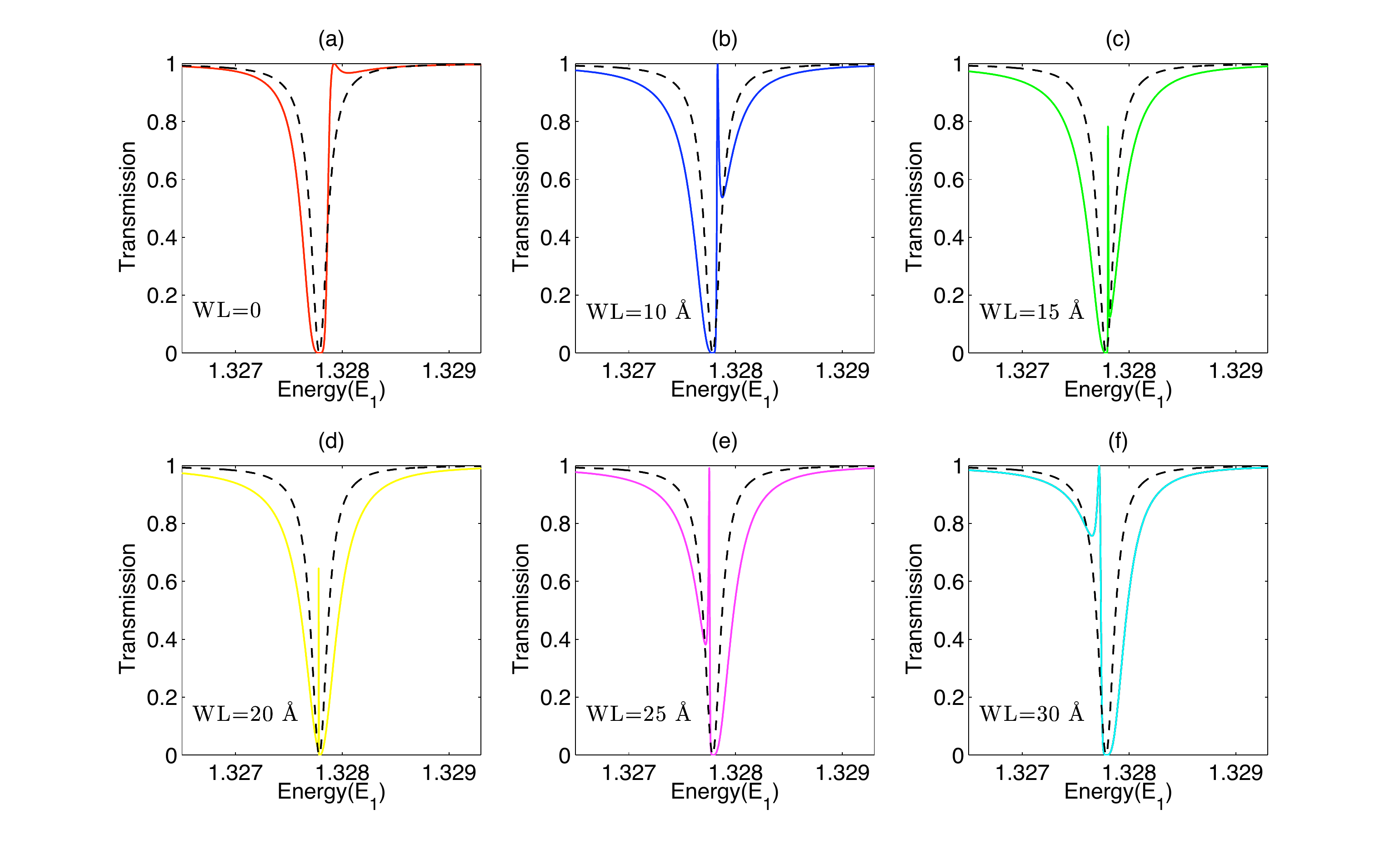}
\end{center}
\caption{\label{fig2} (Color Online) Electron transmission profiles of the lowest resonance for bi-ripple electron waveguides with various distances $WL$ of the flat waveguide between the  two ripple cavities. As $WL$ changes, the width of the two resonances and the resonance energies change. The dashed line is the electron transmission for a  waveguide with a single ripple cavity.}
\end{figure}

Fig.~\ref{fig2} shows the transmission (conductance) of an electron through the bi-ripple waveguide for different flat waveguide lengths  $WL$. The dashed  line is the electron transmission for a  waveguide with a single ripple cavity. We can see the resonance width narrowing and broadening (Dicke effect) depending on $WL$.  

\begin{figure}
\begin{center}
\includegraphics[width=0.5\textwidth]{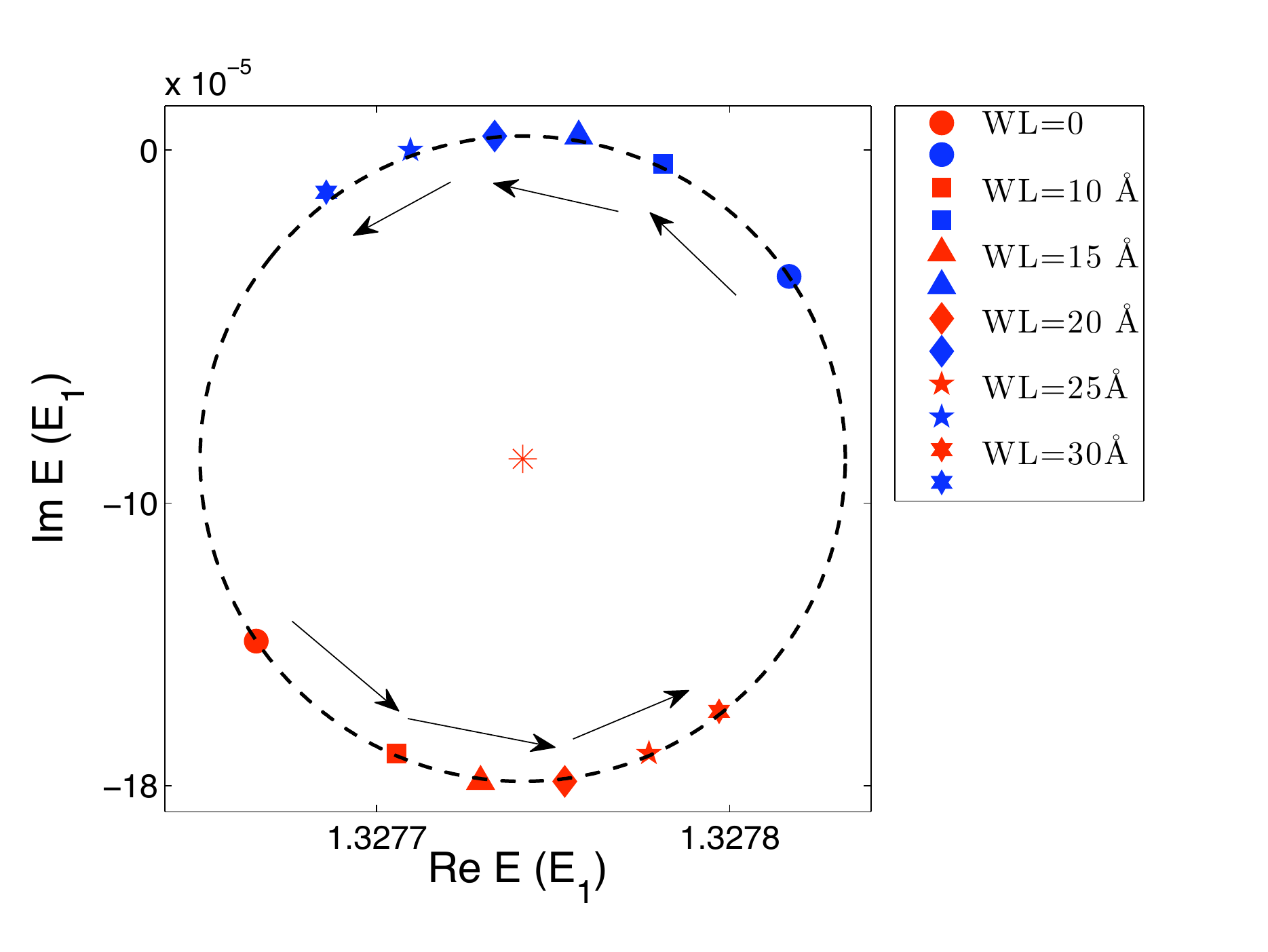}
\end{center}
\caption{\label{fig3} (Color online) S-matrix poles for the bi-ripple waveguide with different values of $WL$. Each pole rotates on a  circle (whose center is the S-matrix pole of the  waveguide(red *) with a single cavity)  in counterclockwise direction as $WL$ increases. The pair of poles has $\pi$ phase difference so that they are located in the opposite direction on the circle. }
\end{figure}

Fig.~\ref{fig3} shows how the S-matrix poles change position  in a complex energy plane. The two poles induce the two resonances in  the transmission plot (Fig.~\ref{fig2}).  The two poles  rotate on a circle (which is centered at the pole position for a waveguide  with a single cavity (red * in Fig.~\ref{fig3}) as the distance $WL$ is varied.  The behavior of the poles indicates that not only the widths but also the resonance energy (real part of the pole) changes with $WL$. The pair of poles is located on opposite sides of the circle,   which means  they have a $\pi$ phase difference. 
As $WL$ increases, one of the poles gets closer to the real axis while the other moves  away  from the real axis (See Fig.~\ref{fig3}). As one of the poles gets closer to the real axis, its corresponding resonance becomes sharper (long-lived state, {\it subradiance}) while the other resonance becomes  broader (short-lived state, {\it superradiance}). The narrowing and broadening of the resonance width is  shown in Fig.~\ref{fig2} (b) $\sim$ (f). 

As the poles rotate, it becomes  possible for one of the poles to reach the real axis and then the width of the resonance collapses, indicating  that the lifetime of the resonance has become  infinite. When this happens, the  resonance state is completely decoupled from external leads. This phenomena has been called ``bound state in continuum" (BIC) in other studies~\cite{Ordonez:2006p1894, Cattapan:2008p2155}.  

\begin{figure}
\begin{center}
\includegraphics[width=0.5\textwidth]{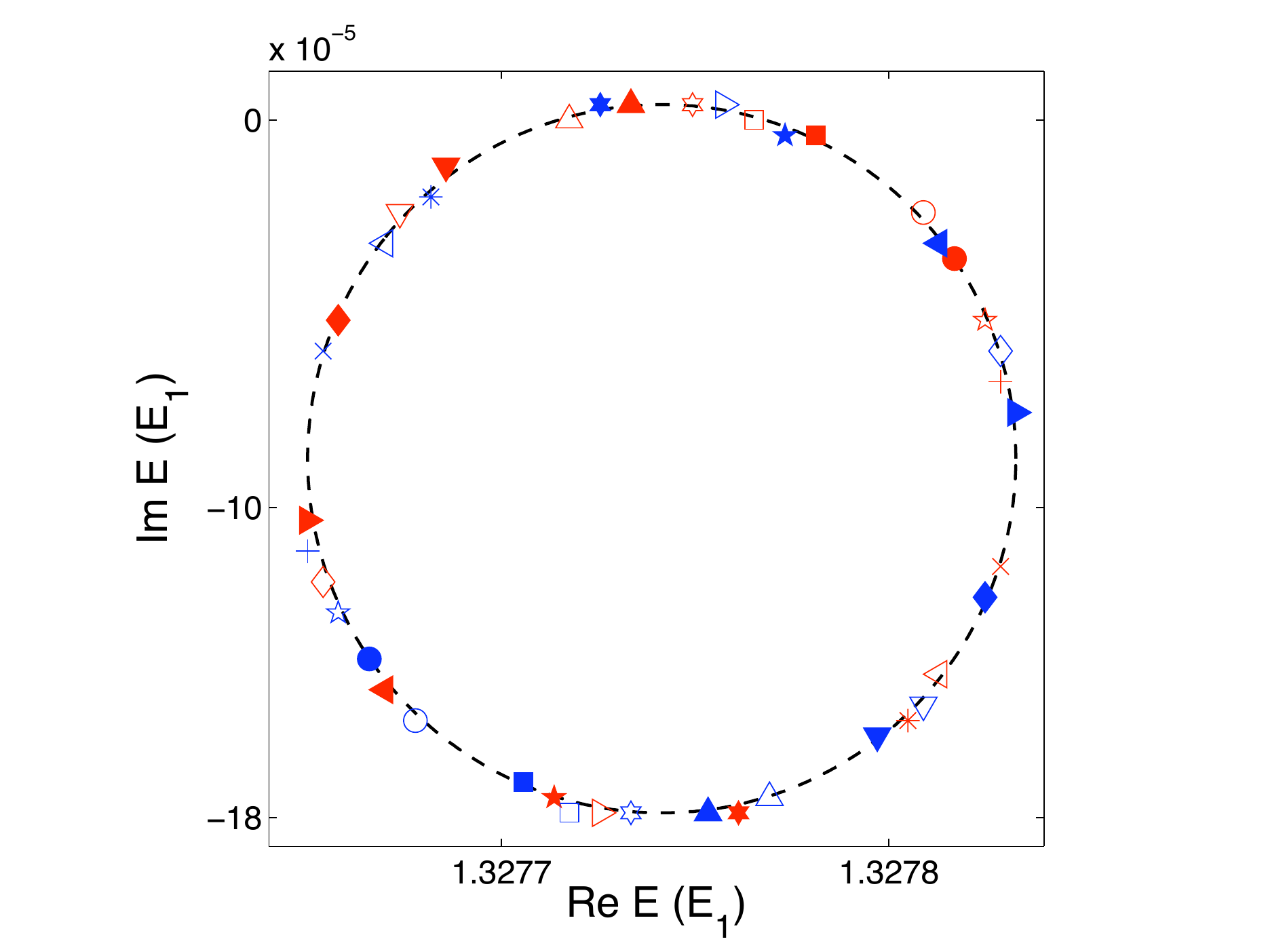}
\end{center}
\caption{\label{fig4} (Color online) S-matrix pole trajectories for bi-ripple waveguide as $WL$ varies from $0$ \AA to $250$ \AA. }
\end{figure}

Let us  express the S-matrix pole position for a  waveguide with a single ripple cavity as  $E =E_0 - i \Gamma_0$ (red * in the Fig.~ref{fig3}).
Since S-matrix poles for the bi-ripple waveguide rotate on a circle in the complex energy plane, the position of the poles can be written as  a sinusoidal function of $WL$ such that,
\begin{equation}
E_1=E_{+} -i {\Gamma}_{+} ~~~{\rm and}~~~E_2= E_{-}- i \Gamma_{-}
\end{equation}
where
\begin{eqnarray}\label{energy}
E_+ &=& E_0 + \Gamma_0 {\rm sin} (kL +\delta_0)  \\
E_- &=& E_0 - \Gamma_0  {\rm sin} (kL +\delta_0) \nonumber
\end{eqnarray}
and
\begin{eqnarray}\label{gamma}
\Gamma_+ &=& \Gamma_0 +\Gamma_0 {\rm cos} (kL + \delta_0) \,=\, \Gamma_0(1+\alpha)\\
\Gamma_- &=& \Gamma_0 -\Gamma_0 {\rm cos} (kL + \delta_0) \,=\, \Gamma_0(1-\alpha) \nonumber .
\end{eqnarray}
The quantity  $\alpha = {\rm cos} (kL + \delta_0)$ is a measure of the coupling between the cavities,  $k$ is a wave number at a resonance energy,  and a phase factor, $\delta_0$ is used because a pole is not on the real axis when $WL=0$. Fig.~\ref{fig4} shows the pole trajectories in the complex energy plane as $WL$ increases up to 250 \AA. It confirms that the  pair of poles always stays on a circle and it verifies Eq.~\ref{energy} and ~\ref{gamma}. 

\begin{figure}
\begin{center}
\includegraphics[width=0.5\textwidth]{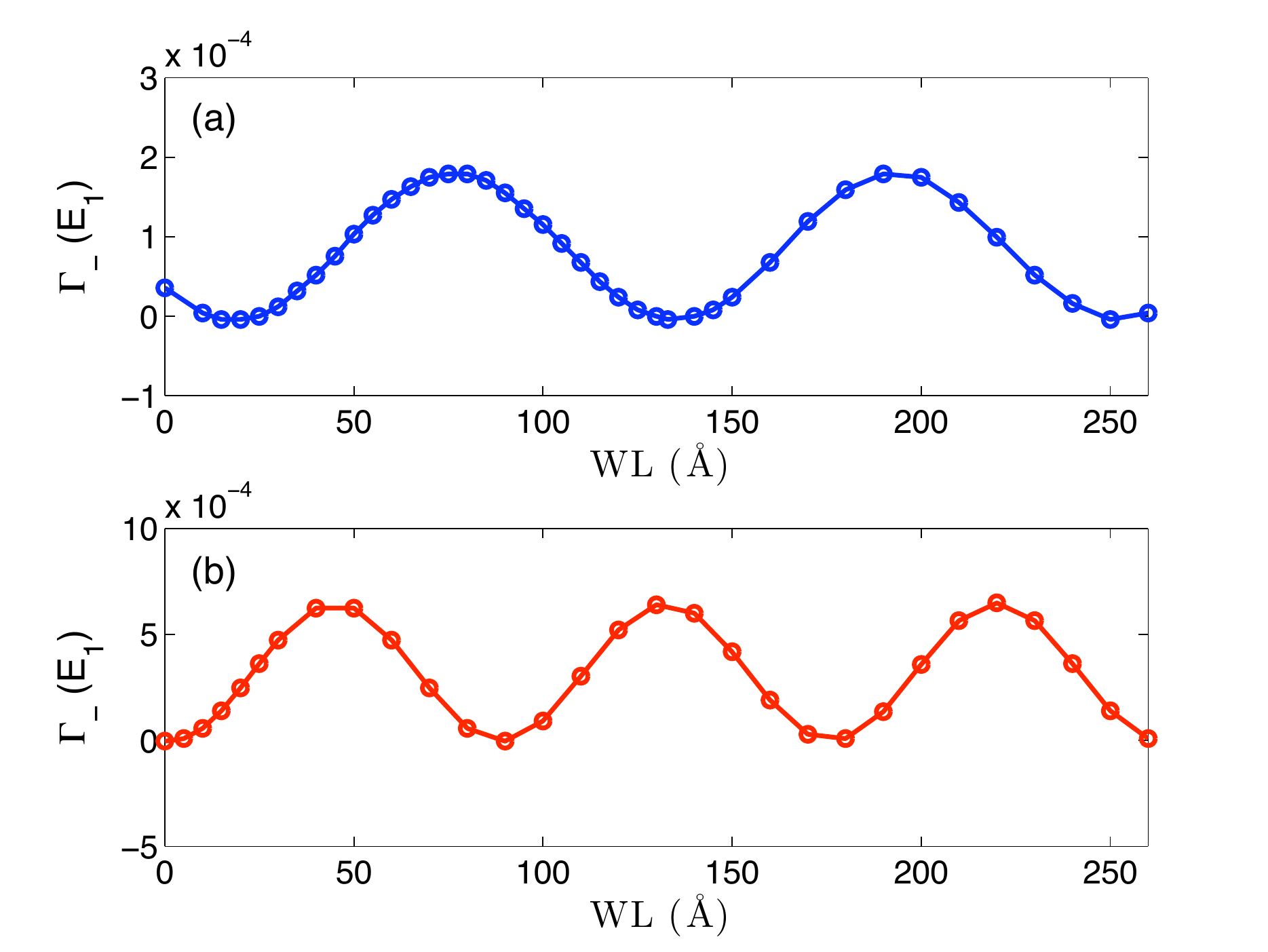}
\end{center}
\caption{\label{fig5} (Color online) The resonance width $\Gamma_-$ for a bi-ripple waveguide. $\Gamma_-$ oscillates as $WL$ is varied. The blue line (a) is for the resonance near $E=1.3278 E_1$.  The red line (b) is for the resonance  near $E=1.5830E_1$. }
\end{figure}

As was shown in Refs.~\cite{Lee:2006p1023} and ~\cite{Lee:2008p840},  a sequence of resonances occurs as energy in the interval $E_1{\leq}E{\leq}4E_1$ is increased. 
Fig.~\ref{fig5} (a)  shows the width of the narrow resonance as a function of the distance $WL$ for the  resonance at energy $E=1.3278E_1$ and \ref{fig5} (b)  shows the width of the resonance at energy $E=1.5830E_1$. The resonance  widths are  clearly  sinusoidal functions of $WL$,  which indicates that  the coupling between the two  cavities  varies  sinusoidally with increasing $WL$. Fig.~\ref{fig5} (b) oscillates slightly faster than \ref{fig5}(a) because of its  higher resonance energy. In Eq.~\ref{gamma}, the coupling parameter $\alpha$ is a sinusoidal function which gives the oscillatory behavior of the width in Fig.~\ref{fig5}. This behavior of the coupling parameter in our 1D system is different from that seen in higher dimension. In Ref.~\cite{Shahbazyan:1994p1192}, the coupling between a collective state of two impurities and external electron wavefunction (a 2D system)  is a Bessel funtion ($J_0(s_{12}/\lambda_f)$) (See Eq. (19) in Ref.~\cite{Shahbazyan:1994p1192}).  In the double quantum dots coupled through the common phonon fields~\cite{Brandes:1999p1470} (a 3D system), the coupling parameter is a zeroth order Bessel function (sin (Qd)/ Qd).  In these 2D and 3D systems, the coupling parameter decays with  the distance between two atoms, as is the case for the Dicke system (a 3D system).  This dependence of the coupling on zeroth order Bessel functions (which have their largest  values  at long wavelength),   means that  in 2D and 3D systems  the superradiance occurs  predominantly with the wavelengths  much longer than the atomic distance.  However, in our 1D waveguide system (the transverse direction is constrained), the coupling parameter ($\alpha$) does not decay.  Therefore,  the superradiant resonance exists regardless of the wavelength of the electron. 
 It is possible that a superradiant resonance appears with a very large distance between two cavities, but it could not exceed the coherence length of electrons in the  waveguide. In 2DEGs  made of GaAs/AlGaAs, for example,  at a very low temperature the coherence length  is about $10 \mu m$. 

As we have seen, the two S-matrix poles induce  two resonances in the electron transmission.  The behavior of  these poles is associated with the symmetry of the scattering wavefunction at resonance energies. In Dicke's model,  {\it subradiant resonance} is related to the anti-symmetric collective state  of the atomic system and the {\it superradiant resonance} is related to the symmetric collective state.  In the  bi-ripple waveguide, as the pair of poles rotate  in a  counterclockwise direction (with increasing distance between the two ripple cavities) , they each maintain the symmetry of their corresponding resonance wavefunction. The antisymmetric state is superradiant when  $\alpha<0 $ and the symmetric state is superradiant when $\alpha>0$.   Therefore, superradiant resonances appear in a sinusoidal manner as $WL$ increases and the symmetry of their corresponding scattering wavefunctions alternates between symmetric and antisymmetric with increasing $WL$. 

In conclusion, we have studied  the behavior  of quasi bound states in a bi-ripple waveguide as the distance between the  two ripple cavities changes. We found that the  widths and positions of the resonances associated with the quasi bound state poles change  in a sinusoidal way with variation of the distance between the ripple cavities. In 2D and 3D models of the Dicke effect, the coupling parameter decays  when the distance between two atoms (impurities or quantum dots) is longer than the wavelength of photon or electron.  In the ripple waveguide system, the coupling parameter does not decay because the waveguide is a quasi 1-D system. 
We also studied the trajectories of S-matrix poles  in the complex energy plane. We found that the pair of poles  that  give rise to the super- and subradiant resonances   in the electron transmission,  rotate on a circle centered on the S-matrix pole for the waveguide with a single cavity and are phase shifted by $\pi$ . Therefore, superradiant and subradiant resonances appear in oscillatory manner as the distance between cavities is changed. Furthermore, as the  S-matrix poles rotate, the symmetry of the electron state associated with the superradiant (subradiant) resonance alternates between symmetric and anti-symmetric with increasing distance between the two ripple cavities.

%
%
%
%

The authors  wish to thank the Robert A. Welch Foundation (Grant No. F-1051) for support of this work.  We also  thank to Dr. Kyungsun Na for her advice and useful discussions, and we  thank   the Texas Advanced Computing Center (TACC) for use of their facilities.


\end{document}